\documentclass[conference]{IEEEtran}
\IEEEoverridecommandlockouts
\usepackage{preamble}
\usepackage[normalem]{ulem}
\usepackage[binary-units]{siunitx}

\def\BibTeX{{\rm B\kern-.05em{\sc i\kern-.025em b}\kern-.08em
    T\kern-.1667em\lower.7ex\hbox{E}\kern-.125emX}}
\begin{document}


\title{Mission-Critical Public Safety Networking: An Intent-Driven Service Orchestration Perspective}

\author{\IEEEauthorblockN{Kashif Mehmood, David Palma, and Katina Kralevska}
\IEEEauthorblockA{\textit{Department of Information Security and Communication Technology (IIK)} \\
\textit{NTNU\textemdash Norwegian University of Science and Technology, Trondheim, Norway}\\
Email: \{kashif.mehmood, david.palma, katinak\}@ntnu.no}
}
\maketitle

\begin{abstract}
\Gls{ibn} provides a promising approach for managing networks and orchestrating services in \gls{b5g} deployments using modern service-based architectures. \Gls{ps} services form the basis of keeping society functional, owing to the responsiveness and availability throughout the network. The provisioning of these services requires efficient and agile network management techniques with low-overhead and embedded intelligence. \Gls{ibn} incorporates the service subscribers in a model-driven approach to provision different user-centric services. However, it requires domain-specific and contextual processing of intents for abstracted management of network functions. This work proposes an intent definition for \gls{ps} services in \gls{b5g} networks, as well as a processing and orchestration architecture for a \gls{ptt} use case. The simulation results show that \gls{ptt} services adhere to the key performance indicators of access time and mouth-to-ear latency bounded by approximately 250 and 150 milliseconds, respectively, with an additional overhead experienced during the intent processing in the range of 20-40 milliseconds. This validates the premise of \gls{ibn} in providing flexible and scalable management and service orchestration solution for \gls{ps} next generation networks. 
\end{abstract}

\glsresetall

\section{Introduction}
 \Gls{ibn} has been extensively studied in recent years focusing on achieving a policy-based high-level control of networked devices using input from subscribers, service providers and network operators~\cite{irtf-nmrg-ibn-concepts-definitions-02}, \cite{ibn-6g-survey}. \gls{ibn} provides a mapping model for user-centric configuration management of coexisting services in communication networks.
 In addition, \gls{ibn} provides a path towards reducing human administration during the service lifecycle management by directly mapping user requests to service profiles and deployment using standard network functions~\cite{9511519, mehmood2021intentslr}.
 This closed-loop orchestration and management of services and network functions opens a deeper discussion towards the automation of lifecycle using, for example, machine learning methods~\cite{ibn-ml-slicing}.   \par

The \gls{ibn} relevance has increased with the softwarization and open implementations of different network functions and service models~\cite{3GPP-28-805-Rel16}. European Telecommunications Standards Institute launched initiatives highlighting the role of \gls{ibn} including Zero Touch Service and Network Management ~\cite{ETSI-ZSM} and Experiential Network Intelligence ~\cite{ETSI-ENI}.
They concluded that intents can embody the user needs in an abstract manner, and that translation helps in mapping the requirements into different data models to be interpreted by service orchestrators and network controllers. \par
\Gls{ps} networks can benefit from the diverse service-based architecture and ecosystem in \gls{b5g} networks~\cite{psc-5g-mec-mcptt}.
However, the diversity in service and vertical environments poses a coexistence challenge for \gls{ps} communication~\cite{psc-5g-access-usecases}.
These verticals coupled with a diverse set of internal use cases and \glspl{kpi} require focused efforts in provisioning and management by the network operators \cite{psc-5g-survey}.

The cost of keeping a dedicated network infrastructure like \gls{tetra} \cite{tetra-etsi} for the \gls{ps} communication is no longer viable due to competitive services offered by commercial cellular networks.
Hence, this assimilation of \gls{ps} services offers mutual benefits for providers and subscribers.
These services encompass multiple scenarios, such as \gls{ptt}, video and data, location tracking, remote control, monitoring of device and on-body sensors.

The utilization of \gls{ibn} as a disruptive network management and service orchestration technology is still being explored~\cite{ibn-toall}, with \gls{ps} communication as a particularly challenging vertical in next generation networks.
\textit{Volk} \textit{et. al}~\cite{psc-5g-access-usecases} studied the feasibility of \gls{5g} for migratory integration of \gls{ps} communication services from the \gls{tetra} infrastructure with experimental validation that \gls{5g} networks can support these services. \textit{Suomalainen} \textit{et. al}~\cite{psc-5g-survey} focused on secure communication within \gls{ps}, exploring the impact on potential military and highly-sensitive use cases.\par
However, there is a gap between service orchestration and management of network functions and the utilization of user input in an effective manner with user centric high-level policies in \gls{ps} scenarios. This paper explores this gap by analyzing the impact of intents over the lifecycle management of \gls{ptt} services in a user-centric orchestration deployment.\par

\begin{figure*}[h!]
\begin{center}
    \includegraphics[width=\textwidth]{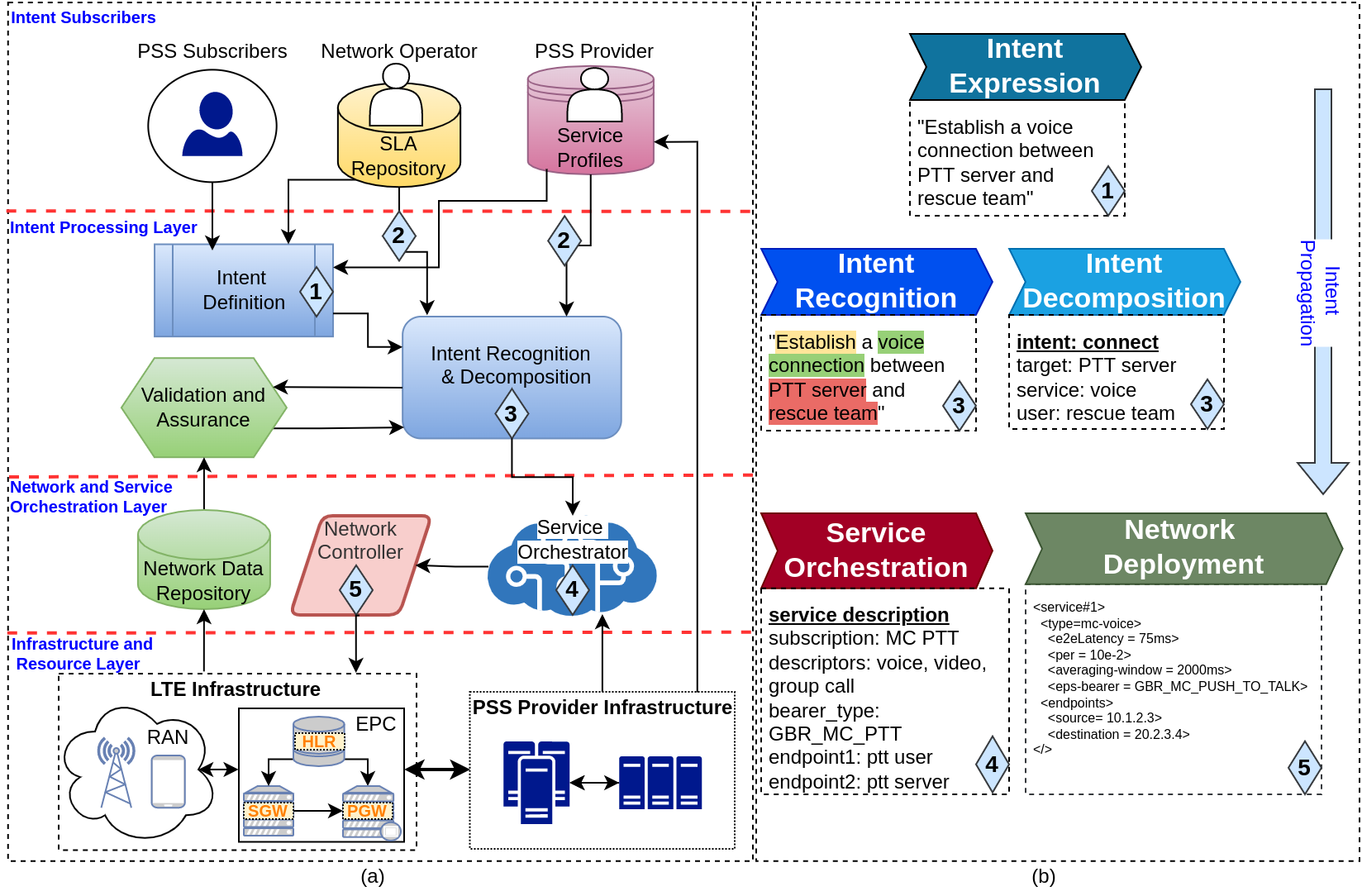}
\end{center}
\caption{(a) Proposed Intent-based network architecture (for \gls{ptt} use case), (b) Intent recognition and translation pipeline. }
\label{image: ibn-arch}
\end{figure*}

We propose the intent-based network management framework, where intents can be generated by a service provider and by a subscriber. Intents are processed and translated into a data model consumed by the network controller through device-level configurations. Resource provisioning is handled through a closed-feedback loop from the underlying LTE infrastructure.
The proposed \gls{ibn} framework takes directly user input by using a \gls{cnl} model, extracting key information related to the service, end points, dependencies, and instructing the service orchestrator to deploy the required resources.
A \gls{ptt} scenario is designed to validate the proposed framework through an extensive performance evaluation of \glspl{kpi}.
The experimental results indicate the ability of different stakeholders to influence the performance of \gls{ps} service in distinct communication scenarios encountered by the subscribers.

The rest of the paper is organized in the following manner. Section II presents the proposed framework for intent-driven network management and service orchestration of  \gls{ps} network. The simulation design methodology and experimental results for the  \gls{ptt} communication scenario are presented in Section III, along with a discussion of key findings. Finally, Section IV concludes this paper.   

\section{Proposed Intent-Based Network Management Framework for PS networks}
\Gls{ps} communication encompasses an ideal scenario for deterministic networking with guaranteed performance for different use cases. \Gls{ptt} is one such use case for rescue operations performed by different health and safety authorities \cite{mcptt-3gpp}. This consists of assisting different public safety authorities in communicating with low-latency and high reliability guarantees. It also supports different users with strict connection setup and latency requirements (Sec-IIB) in order to coordinate an evolving rescue operation. In other words, \gls{ptt} falls under the umbrella of ultra-reliable and low latency communication services making it a challenging service to be provisioned to subscribers using traditional resource allocation methods.  \par
\begin{table*}[]
\centering
\caption{CNL based intent translation and mapping logistics.}
\label{tab: cnl-map}
\begin{tabular}{|c|l|l|l|}
\hline
\textbf{Connectivity}                      & \begin{tabular}[c]{@{}l@{}}'establish', 'connect', 'reconnect',\\  'provide', 'enable'\end{tabular} & \begin{tabular}[c]{@{}l@{}}'disconnect', 'terminate', \\ 'disable', 'tear down'\end{tabular} & \begin{tabular}[c]{@{}l@{}}'mc server', 'mc user', 'ptt server', \\ 'ptt user'\end{tabular} \\ \hline
\textbf{Service}                           & 'voice'                                                                                             & 'video'                                                                                      & 'signaling'                                                                                 \\ \hline
\textbf{Potential LTE Bearer Type} & GBR\_MC\_PUSH\_TO\_TALK                                                                             & GBR\_MC\_VIDEO                                                                               & NGBR\_MC\_DELAY\_SIGNAL                                                                     \\ \hline
\end{tabular}
\end{table*}

Exposure of system-level information and methods towards service subscribers for different communication services is critical towards designing user-centric service and network orchestration policies. However, the unavailability of contextual information (e.g., \gls{sla}, network control policies) for service subscribers poses a challenge. Instead, subscriber intents can be incorporated into the decision making and orchestration of different services and network infrastructure. This is accomplished through an intricate expression and recognition of subscriber input followed by the decomposition, contextual mapping and deployment of user-centric policies.\par

The management flow starts with intent sources, i.e. \gls{ps} service subscribers and providers, acting as the intent generators and specifying high-level abstract policies. The mapping, validation and conversion of intents to low-level service profiles is done and presented to the network controller for deployment. This intent processing pipeline has been labelled as steps \textbf{1} to \textbf{5} in Fig. \ref{image: ibn-arch}. The underlying physical network infrastructure consists of \gls{ptt} service deployed through a dedicated \gls{ps} service infrastructure provider.\par

The service subscriber generates an intent, that is processed with input from the network and service provider resulting in a set of services, connection endpoints and required networking capabilities. The service environment adapts based on the changes in the active service flows as per different intents in the network. For example, a rescue team can prompt the initiation of a group conversation, localized and tracked user monitoring or, in case of going out-of-coverage, a relay connectivity solution for some team members.

\subsection{IBN Model}
The network infrastructure is based on LTE deployment for this study and \gls{ps} service is provided through a service provider with various service offerings. Intents provide a critical integration tool to handle different types of service requests with minimum overhead ensuring reliable service delivery.
The declarative definition of intents shields the service subscribers and providers from resource as well as infrastructure level-state of the network. The onus of understanding these high-level policies falls upon the network operator requiring the need for a dedicated intent-processing and translation component.

\subsubsection{Intent definition and processing}
Recent studies in \gls{ibn} have focused on a contextual intent description model or a domain-specific language (DSL) model to represent intents~\cite{ibn-toall}. This approach provides domain specific intent processing lacking generalization of the intent processing pipeline. In contrast, we specify that intents are defined using a set of \gls{cnl} instructions with the core components shown in Fig. \ref{image: ibn-arch}(b) step \textbf{1}. The service subscriber utilizes a simple CNL model with keywords specified in the Table \ref{tab: cnl-map} to express the desired connection and service in a declarative manner. Intents can be generated using the combination of keywords with desired endpoints and services from Table-I.  In this paper, the proposed framework utilizes the generation of natural language based intent expression model followed by conversational mapping of the required service requirements and orchestration by the orchestrator. The type of intent is specified by the vocabulary of the noun phrases used in intent expression. Moreover, the knowledge of available services offered by the service provider gives the required scope of the intent expression. The network level knowledge database consists of \glspl{sla}, offered services, service providers and network resource capabilities expressed in coordination with infrastructure and deployment information.

\subsubsection{Intent decomposition, recognition, and deployment}
The recognition of the core elements forming the intents provides an initial step towards developing an understanding and eventual mapping operations. The mapping of high-level declarative policies into an orchestrator agnostic template is the prime objective of the intent processing module. The translation is accomplished in close coordination with the network operator, considering the available data sources in the knowledge database (step \textbf{2}). The relevant service keyword is identified from the intent and mapped onto a supported service profile from the available set of service offerings in the network (step \textbf{3}). The mapped service is prepared for deployment through the construction of service descriptors with information extracted from the underlying service provider infrastructure to create service deployment templates (step \textbf{4}). A network controller then constructs the LTE bearer configuration template using SLOs to be deployed in the network infrastructure completing the orchestration of recognized services  (step \textbf{5}).

\subsubsection{Validation and Assurance}
The validation and feasibility analysis of the intent is performed with the help of telemetry data from the network devices. A dialogue can be sought with the intent user in case of failure to verify the sufficiency of resources, possible conflicts with deployed intents or lack of required subscription with the service provider. The orchestration is initiated upon successful completion of the intent validation. Moreover, deployed intent performance is monitored through collected network data and compliance with \gls{sla} is ensured through the intent processing modules.
\begin{figure*}[h!]
\includegraphics[width=\textwidth]{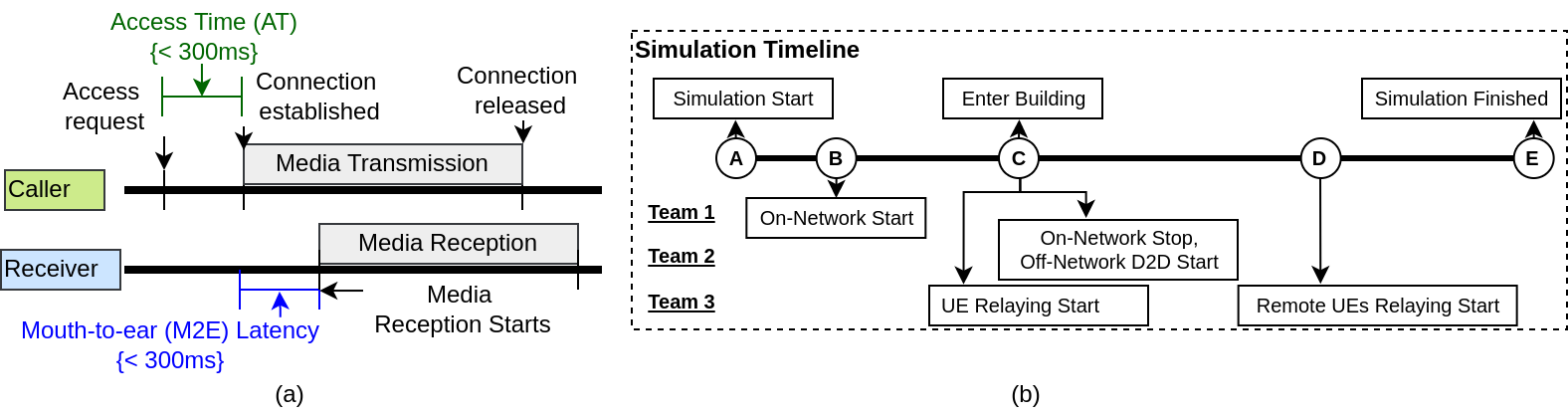}
\caption{(a) \gls{ptt} \glspl{kpi}, (b) Key events during simulated intent-based \gls{ptt} scenario.}
\label{image: ptt-kpi}
\end{figure*}

\subsection{PS Network Infrastructure}
\subsubsection{\gls{ps} Network Model}
In this paper, we consider an LTE network environment where the intent based \gls{ptt} service orchestration is performed. Three different teams with varying number of user devices are coordinating a rescue operation inside a building and devices in a team can connect to one or multiple devices simultaneously based on the nature of the rescue operation. The radio access network supports on-network, off-network (\gls{d2d}) and \gls{ue} relays in order to maintain connection amongst different rescue teams based on \gls{enb} coverage. The orchestrated \gls{ptt} service provides the user devices in different teams to choose from the available modes of connectivity throughout the course of the rescue operation.

\subsubsection{\gls{ps} Service Use Case}
\Gls{ptt}, video and data connections form the basis of the considered system model, however it can be extended if needed, given that the \gls{ibn} model uses an abstract policy definition with mapping of \glspl{kpi}. \gls{ps} services have stringent \gls{qos} constraints made specified by the \gls{3gpp} from Release-12 up til Release-17~\cite{5g-3gpp}.\par
\Gls{ptt} supports one-to-one, group calls, and emergency calls, where the caller first requests the resources, and the grant is provided by the network in coordination with the \gls{ptt} service provider, followed by  call termination. The connection is established in a sequence of events:
\begin{itemize}
    \item A user requests the floor to access the network in order to utilize the \gls{ptt} service (access request in Fig. \ref{image: ptt-kpi}(a)).
    \item The network operator grants the floor to the requesting user based on resource availability.
    \item Afterwards, the user can initiate the data transfer to one or multiple receivers (media transmission in Fig. \ref{image: ptt-kpi}(a)).
\end{itemize}

\subsubsection{PSS \glspl{kpi}}
\Gls{ptt} is a typical use case of \gls{ps} service. \gls{3gpp} considers several components of access and communication delay coupled with propagation delays through the access and core infrastructure to model the following \glspl{kpi}~\cite{mcptt-3gpp}: 
\begin{itemize}[noitemsep,topsep=2pt,parsep=0pt,partopsep=0pt]
    \item \textit{\Gls{at}}: It is the time between user request and grant of permission to speak.
    \item \textit{\gls{m2e} latency}: It is the time between an instance when sender speaks and the instance when receiver hears the transmission.
\end{itemize}
The timeline for the measurement of these \glspl{kpi} as well as the performance requirements specified by 3GPP Release 16 \cite{5g-3gpp}, are shown in \figurename~\ref{image: ptt-kpi}(a).

\section{Performance Evaluation}
In this section, we describe the simulation design for the validation of the proposed framework for an intent-based provisioning of \gls{ps} services. We evaluate the performance of a \gls{ptt} scenario, with the help of \gls{at} and \gls{m2e} latency as relevant \glspl{kpi}. The simulation setup is designed to mimic the behavior of a \gls{ptt} provides the base infrastructure for deployment and assurance of  intents through relevant service requirements \cite{tetra-etsi}.\par
Key simulation parameters are provided in Table \ref{table: sim-params} for the \gls{ptt} scenario in Fig. \ref{image: ptt-kpi} (b). The figure describes the sequence of events for different teams in the simulation environment in order to produce a varied performance behavior. This enables quantification of performance of a \gls{ptt} scenario once the deployment has been done using IBN framework. 
\begin{table}[]
\caption{A list of simulation parameters.}
\label{table: sim-params}
\begin{tabular}{|lll|}
\hline
\multicolumn{3}{|c|}{\textbf{Simulation Parameters}}                                                                                                                                                    \\ \hline
\multicolumn{2}{|l|}{\# of simulation runs}                                                                        & 20                                                               \\ \hline
\multicolumn{2}{|l|}{Network Mode}                                                                                 & \begin{tabular}[c]{@{}l@{}}On (Team 1), Off (Team 2)\\ Relay (Team 3)\end{tabular} \\ \hline
\multicolumn{2}{|l|}{\# of users\textbackslash{}team}                                                              & {[}4, 8, 12, 16{]}                                                                 \\ \hline
\multicolumn{1}{|l|}{\multirow{3}{*}{Distance}}        & \multicolumn{1}{l|}{eNB\textless{}-\textgreater{}Ue}      & 400 m                                                                              \\ \cline{2-3} 
\multicolumn{1}{|l|}{}                                 & \multicolumn{1}{l|}{Ue\textless{}-\textgreater{}building} & 10 m                                                                               \\ \cline{2-3} 
\multicolumn{1}{|l|}{}                                 & \multicolumn{1}{l|}{Ue\textless{}-\textgreater{}Ue}       & 5 m                                                                                \\ \hline
\multicolumn{2}{|l|}{PTT floor queueing}                                                                           & {[}on, off{]}                                                                      \\ \hline
\multicolumn{1}{|l|}{\multirow{2}{*}{Transmit Power}}  & \multicolumn{1}{l|}{Ue}                                   & {[}23, 25, 27{]} dBm                                                               \\ \cline{2-3} 
\multicolumn{1}{|l|}{}                                 & \multicolumn{1}{l|}{eNB}                                  & {[}30, 32, 34{]} dBm                                                               \\ \hline
\multicolumn{1}{|l|}{\multirow{2}{*}{Pathloss Models}} & \multicolumn{1}{l|}{Outdoor}                              & Hata, COST231                                                                      \\ \cline{2-3} 
\multicolumn{1}{|l|}{}                                 & \multicolumn{1}{l|}{Indoor}                               & ITU-R P.1411 , ITU-R P.1238                                                        \\ \hline
\multicolumn{1}{|l|}{LTE Sidelink}                     & \multicolumn{1}{l|}{MCS}                                  & 15                                                                                 \\ \hline
\multicolumn{2}{|l|}{\# of RBs per UE}                                                                             & 5                                                                                  \\ \hline
\multicolumn{2}{|l|}{Scheduler}                                                                                    & MinProb                                                                            \\ \hline
\multicolumn{2}{|l|}{RSRP Threshold (Relaying)}                                                                    & -124 dBm                                                                           \\ \hline
\end{tabular}
\end{table}
\begin{figure}[t!]
\includegraphics[width=0.48\textwidth]{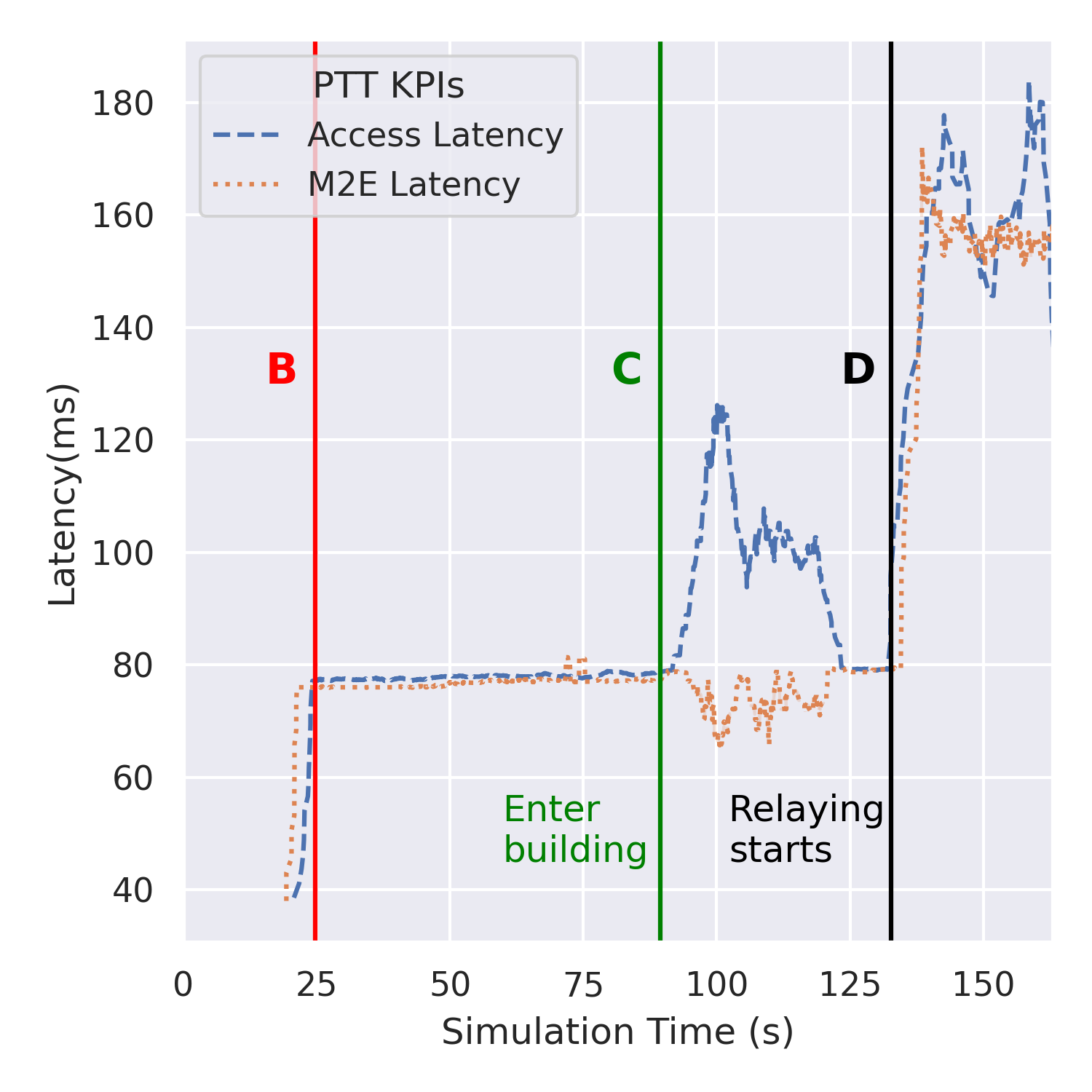}
\caption{A timeline of KPI performance with simulation time.}
\label{image: sim-time}
\end{figure}

\begin{figure*}[ht!]
	\begin{subfigure}[b]{0.49\textwidth}
		\includegraphics[width=1\textwidth]{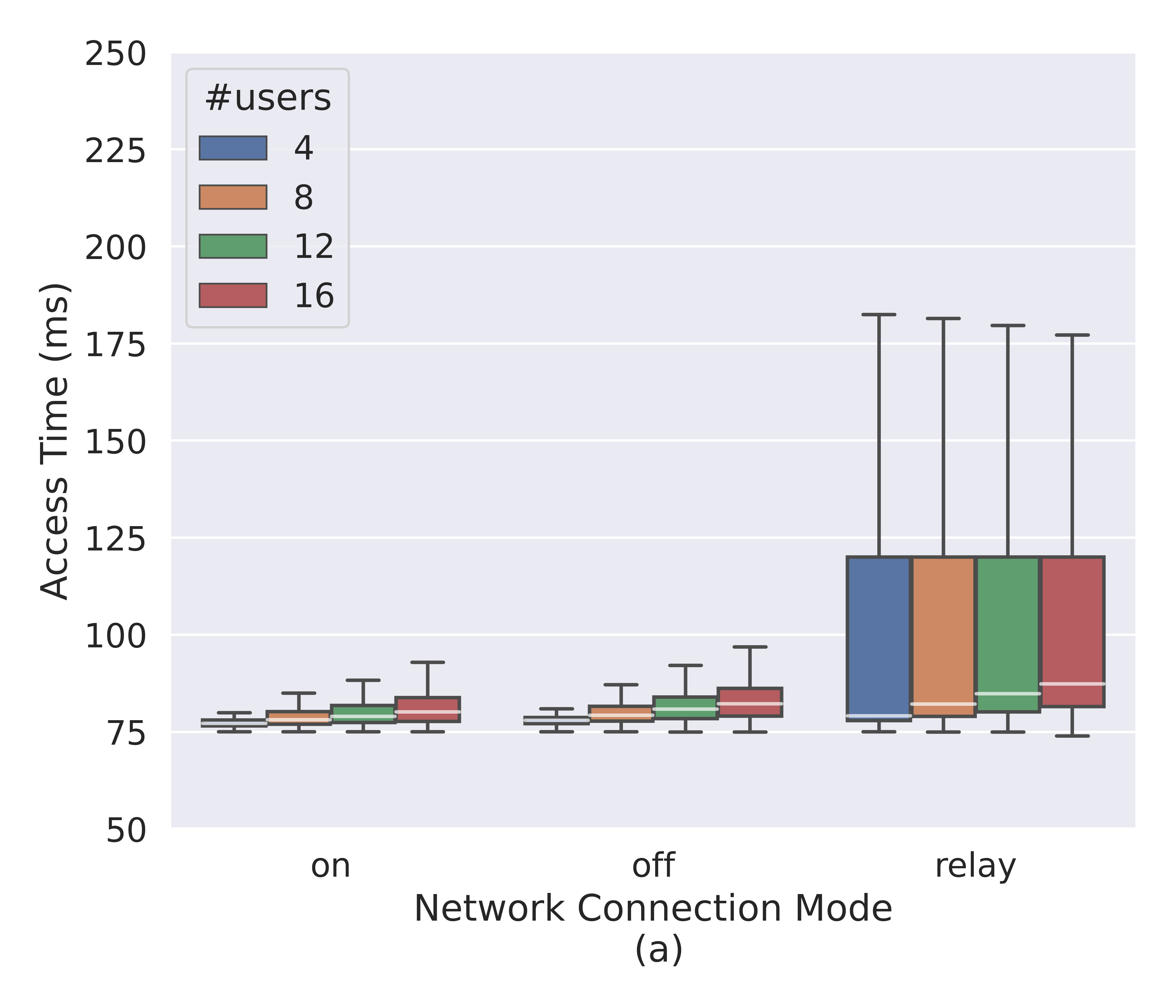}
	\end{subfigure}
	\begin{subfigure}[b]{0.49\textwidth}
		\includegraphics[width=1\textwidth,]{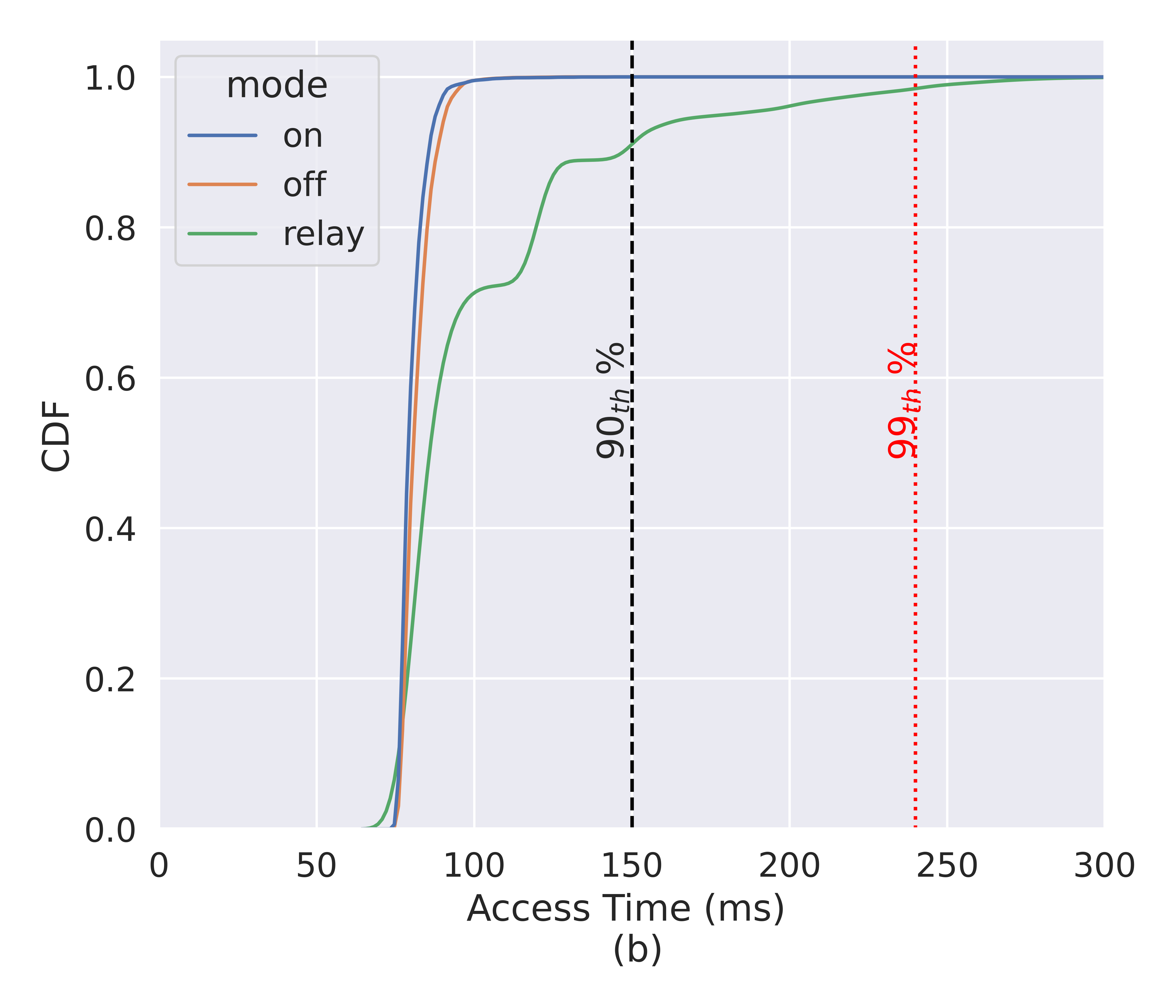}
	\end{subfigure}
	\caption{Access Time (AT) performance with respect to number of users per team and communication modes for 20 runs.}
	\label{image:kpi1}
\end{figure*}
\subsection{Experimental Setup}
The simulation involves two parts -- intent processing and service orchestration. Intents are processed through a \gls{cnl} model in Python interacting with the service orchestrator implemented using \gls{ns-3} \cite{ns3} . The simulation scenario along with the sequence of events triggered by the intents is shown in \figurename~\ref{image: ptt-kpi} (b). The connection intents are generated by the users when the network simulation starts at instant \textbf{A} for service orchestration by the orchestrator followed up by necessary resource allocation.

The \gls{ptt} network consists of a single \gls{enb} connected to an LTE enhanced packet core (EPC). A total of three rescue teams consisting of varying number of \glspl{ue} are available in the scenario for utilizing the \gls{ptt} service. One \gls{ue} amongst each team acts as an anchor point with no mobility, while, multiple UEs within a team move inside the building to their respective locations.\par

The simulation can essentially be divided into three parts consisting of on-network, off-network and relay mode for different \glspl{ue}.The following events occur as shown in the timeline in Fig. \ref{image: sim-time}: Team 1 is provided on-network access through the \gls{enb} and starts a call at instant \textbf{B}. Team 2 stops its in-network access upon reaching the building (instant \textbf{C})  and starts a group-call in off-network mode using \gls{d2d} connection. Team 3 connects normally to the \gls{enb}, starts a call at instant \textbf{B}, but switches to its stationary anchor \gls{ue} for relaying as soon as the received power from the \gls{enb} falls below a threshold. The users continue to move inside the building while connected to their respective network modes until instant \textbf{E} is reached.

\begin{figure*}[ht!]
	\begin{subfigure}[t]{0.49\textwidth}
		\includegraphics[width=1\textwidth]{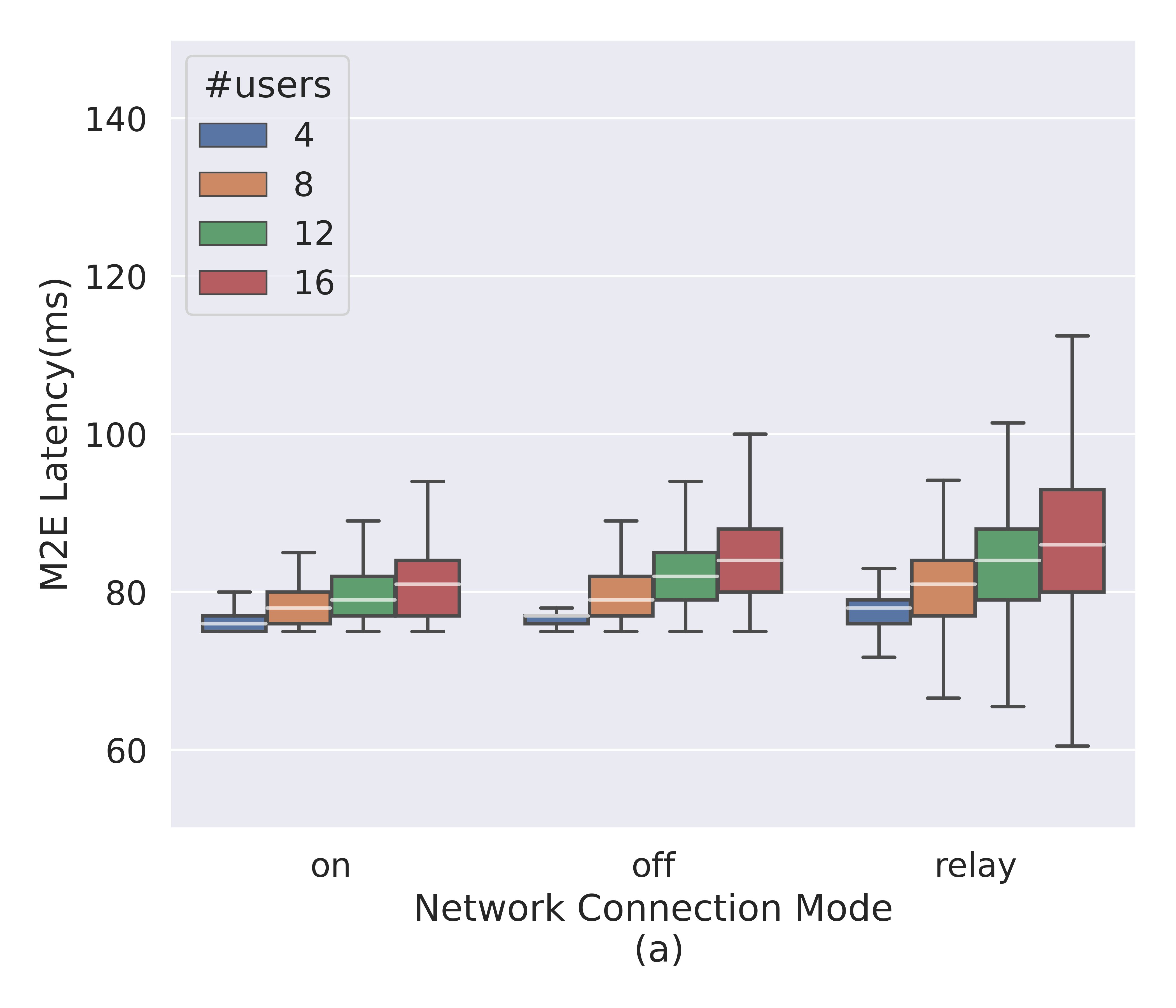}
	\end{subfigure}
	\begin{subfigure}[t]{0.49\textwidth}
		\includegraphics[width=1\textwidth,]{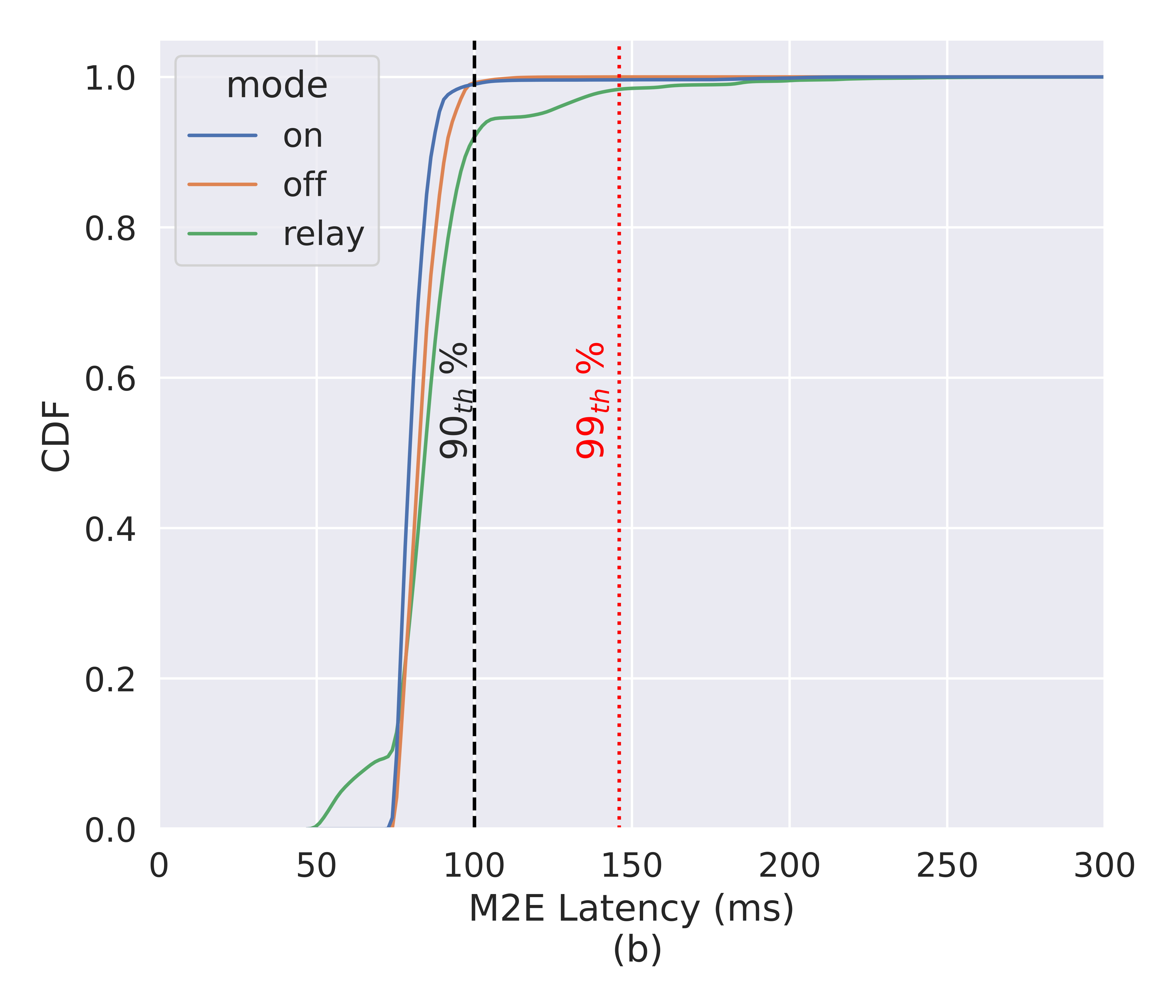}
	\end{subfigure}
	\caption{M2E Latency performance with respect to number of users per team and communication modes for 20 runs.}
	\label{image:kpi3}
\end{figure*}

\subsection{Experimental Results and Discussion}
The simulations provide a contextual overview of the \gls{ptt} use case with varying levels of traffic load and the need for adaptive network connectivity during a \gls{ps} operation. We analyze the performance of the \gls{ptt} scenario based on access time (\gls{at}) and \gls{m2e} latency.
\subsubsection{\gls{ibn} performance}
The \gls{cnl} model decomposes and maps the intent components to service profiles and simulation parameters to be utilized by \gls{ns-3}. The service type, end points and dependencies are translated into specific data models for the given service. For example, the \gls{ptt} service provider has some mappings for the servers handling the information related to the \gls{ptt} requests such as emergency call, group call, and one-to-one call, which are given in the form of service profiles to the intent processing model. The intent translation time varies between 20ms and 40ms depending on the type of intent defined during the processing phase, regardless of the service-specific parameters. This is orders of magnitude lower than the values observed for the access time for the \gls{ptt} service.

\subsubsection{\gls{ps} network and service performance}
The simulation architecture provides the device and resource level exposure for the orchestration of \gls{ptt} services to different rescue teams. After the \gls{ptt} users generate the connectivity intent, the \gls{ps} service provider now acts as an intent user and forwards the connectivity intent to acquire provisioning resources for different teams. The established connection provides the ability to switch amongst different network modes for providing the required access to all the teams in the network. Access time and \gls{m2e} latency are measured for different users as they connect to the network using different modes with results depicted in Fig. \ref{image:kpi1} and Fig. \ref{image:kpi3}, respectively.\par

Access time performance is visualized for different network modes and varying \gls{ptt} user densities per team in Fig. \ref{image:kpi1} (a) with accompanying cumulative distribution function (CDF) in Fig. \ref{image:kpi1} (b). It is observed that on-network mode for team 1 performs the best, followed by the users from team 2 switching to off-network \gls{d2d} mode as they move inside the building. Relaying-based access available to team 3 users performs the worst due to the bottleneck at the relaying UE causing significant variance. The performance of the off-network and relaying mode is also impacted by the change network mode. Fig. \ref{image:kpi1} (b) shows that 99 percent of users have an access latency bound of approximately 250ms and rare violations above it.  The deviation from the observed range of \gls{at} values increases with the density of \gls{ptt} users per team due to queuing before the floor is granted.\par

\gls{m2e} latency is observed to be within the range of the threshold defined by 3GPP~\cite{5g-3gpp} as shown in Fig. \ref{image:kpi3} (a). It is observed that on- and off-network M2E latency is stable with little deviations mainly attributed to the varying simulation attributes. In addition, the observed performance in relaying mode is mainly due to the transmission bottleneck from team 3 anchor node and the increased user density in the access network. Fig. \ref{image:kpi3} (b) shows that 90 percent of \gls{ptt} users experience M2E latencies of 100ms or less and very few exceptions above 200ms. The saturation observed in M2E latency values for a higher number of users is directly proportional to the communication mode and the available resources in the simulation environment.

\section*{Conclusion}
This paper studied the potential application of \gls{ibn} in a \gls{ps} \gls{ptt} scenario resulting in provisioning and modification of services during the service lifecycle. \gls{ibn} provides the basis for a flexible, scalable and service agnostic design of a network management solution. The service subscribers and providers define high-level abstract connectivity and reconnection intents, that are processed by the proposed framework. The service information is extracted from the intents and mapped onto low-level configurations deployed through a network controller to a public safety network. Simulation results show the possibility of incorporating user intents in the orchestration of services in coordination with different stakeholders with minimum impact on the access and M2E latency.


\bibliographystyle{ieeetr}
\typeout{}
\bibliography{refs}

\end{document}